\documentclass{article}%
\usepackage{amsmath}
\usepackage{amsfonts}
\usepackage{amssymb}
\usepackage{graphicx}%
\begin{document}

\title{Maximum Entropy and Bayesian Data Analysis: Entropic Priors}
\author{Ariel Caticha$^{*}$ and Roland Preuss$^{\dagger}$\\$^{\ast}${\small Department of Physics, University at Albany-SUNY, }\\{\small Albany, NY 12222, USA.}\\$^{\dagger}${\small Center for Interdisciplinary Plasma Science, }\\{\small Max-Planck-Institut f\"{u}r Plasmaphysik, EURATOM Association, }\\{\small Boltzmannstrasse 2, D-85748 Garching bei M\"{u}nchen, Germany}}
\date{}
\maketitle

\begin{abstract}
The problem of assigning probability distributions which objectively reflect
the prior information available about experiments is one of the major
stumbling blocks in the use of Bayesian methods of data analysis. In this
paper the method of Maximum (relative) Entropy (ME) is used to translate the
information contained in the known form of the likelihood into a prior
distribution for Bayesian inference. The argument is inspired and guided by
intuition gained from the successful use of ME methods in statistical
mechanics. For experiments that cannot be repeated the resulting
\textquotedblleft entropic prior\textquotedblright\ is formally identical with
the Einstein fluctuation formula. For repeatable experiments, however, the
expected value of the entropy of the likelihood turns out to be relevant
information that must be included in the analysis. The important case of a
Gaussian likelihood is treated in detail.

\end{abstract}

\section{Introduction}

The inference of physical quantities from data generated either by experiment
or by numerical simulation is a ubiquitous and often cumbersome task. Whether
the data is corrupted by noise, hampered by finite resolution or tied up in
correlations, in principle it should always be possible to improve the
analysis by taking into account, in addition to the information contained in
the data, whatever other knowledge one might have about the physical
quantities to be inferred or about how the data was generated. The way to link
this prior information with the new information in the data is found in
Bayesian probability theory.

Bayesian methods are increasingly popular in physics \cite{Dose03}. They are
essential whenever repeating the experiment many times in order to reduce the
measurement uncertainty is either too expensive or time consuming. This is a
common situation in astronomy and astrophysics \cite{Feigelson92}, and also in
large laboratory experiments as in fusion \cite{Dose98} and in high energy
physics \cite{DAgostini99}. Other typical uses in physics arise in spectrum
restoration, in ill-posed inversion problems
\cite{vdLinden93,Preuss94,Fischer97} and when separating a signal from an
unknown background \cite{vdLinden99}. Applications include mass spectrometry
\cite{Schwarz01}, Rutherford backscattering \cite{vToussaint99} and nuclear
magnetic resonance \cite{Bretthorst88}.

From a general point of view the problem of inductive inference is to update
from a prior probability distribution to a posterior distribution when new
information becomes available. The challenge is to develop updating methods
that are systematic and objective. Two methods have been found which are of
very broad applicability: one is based on Bayes' theorem and the other is
based on the maximization of entropy. The choice between these two updating
methods is dictated by the nature of the information being processed.

When we want to update our beliefs about the values of certain quantities
$\theta$ on the basis of the observed values of other quantities $y$ -- the
data -- and of some known relation between $\theta$ and $y$ we must use Bayes'
theorem. The updated or posterior distribution is $p(\theta|y)\propto
\pi(\theta)p(y|\theta)$; the relation between $y$ and $\theta$ is supplied by
a known model $p(y|\theta)$; the previous knowledge about $\theta$ is codified
both into the \textquotedblleft prior\textquotedblright\ probability
$\pi(\theta)$ and also in the \textquotedblleft likelihood\textquotedblright%
\ distribution $p(y|\theta)$.

The selection of the prior $\pi(\theta)$ is a controversial issue which has
generated an enormous literature \cite{Kass96}. The difficulty is that it is
not clear how to carry out an objective translation of our previous beliefs
about $\theta$ into a distribution $\pi(\theta)$. One reasonable attitude is
to admit subjectivity and recognize that different individuals may start from
the same information and legitimately end with different translations. In
simple cases experience and physical intuition have led to a considerable
measure of success, but we are often confronted with new complex situations
involving perhaps parameter spaces of high dimensionality where we have
neither a previous experience nor a reliable intuition.

On the other hand, there are special cases where some degree of objectivity
can be attained. For example, requirements of invariance can go a long way
towards the complete specification of a prior. Considerable effort has been
spent seeking an objective characterization of that elusive state of knowledge
that presumably reflects complete ignorance. Although there are convincing
arguments against the existence of such non-informative priors
\cite{Bernardo97}, the search has had the merit of suggesting connections with
the notion of entropy \cite{Jaynes68} including two proposals for
\textquotedblleft entropic priors\textquotedblright\ \cite{Skilling89,
Rodriguez89}. This brings us to the second method of processing information.

Bayes' theorem follows from the product rule for joint probabilities,
$p(y,\theta)=\pi(\theta)p(y|\theta)$, and therefore its applicability is
restricted to situations where assertions concerning the joint values of the
data $y$ and the parameters $\theta$ are meaningful. But there are situations
where the available information is of a different nature and involves
assertions about the probabilities themselves. Such information, which
includes but is not limited to assertions about expected values, cannot be
processed using Bayes' theorem.

The method of Maximum Entropy (ME) is designed for updating from a prior
probability distribution to a posterior distribution when the information to
be processed takes the form of a constraint on the family of acceptable
posterior distributions \cite{footnote1}. The early and less satisfactory
justification of the ME method followed from interpreting entropy, through the
Shannon axioms, as a measure of the amount of uncertainty in a probability
distribution \cite{Shannon48, Jaynes57}. Objections to this approach are that
the Shannon axioms refer to probabilities of discrete variables, the entropy
of continuous variables is not defined, and that the use of entropy as the
unique measure of uncertainty remained questionable. Other so-called entropies
could and, indeed, were introduced. Ultimately, the real problem is that
Shannon was not concerned with inductive inference. He was not trying to
update probability distributions but was instead analyzing the capacity of
communication channels. Shannon's entropy makes no reference to prior distributions.

Considerations such as these motivated several attempts to justify the ME
method directly as a method of inductive inference without invoking
questionable measures of uncertainty \cite{ShoreJohnson80, Skilling88}. The
concept of relative entropy is then introduced as a tool for consistent
reasoning which, in the special case of uniform priors, reduces to the usual
entropy. There is no need for an interpretation in terms of heat, disorder, or
uncertainty, or even in terms of an amount of information. Perhaps this is the
explanation of why the search for the meaning of entropy has turned out to be
so elusive: strictly, \emph{entropy needs no interpretation}. In section 2, as
background for the rest of the paper, we present a brief outline of one such
`no-interpretation' approach inspired by \cite{Skilling88}.

In this paper we use entropic arguments to translate prior information into a
prior distribution. Rather than seeking a totally non-informative prior, we
make use of information that we do in fact have. Remarkably, it turns out that
the very conditions that allow us to contemplate using Bayes' theorem --
namely, knowledge of a likelihood function, $p(y|\theta)$ -- already
constitute valuable prior information. In this sense one can assert that the
search for completely non-informative priors is misplaced: if we do not know
the likelihood, then prior distributions are not needed anyway. The prior thus
obtained is an \textquotedblleft entropic prior.\textquotedblright\ The name
and the first proposal of a prior of this kind is due to Skilling
\cite{Skilling89} for the case of discrete distributions. The generalization
to the continuous case and further elaborations by Rodr\'{\i}guez
\cite{Rodriguez89, Rodriguez02} constitute a second proposal.

It is essential for the successful use of any prior, and of entropic priors in
particular, to be aware of what information they contain and, crucially, what
information they do not contain. No prior can be expected to succeed unless
all the information relevant to the problem at hand has been taken into
account. It is quite likely that most practical problems that were encountered
with entropic priors in the past can be traced to a failure to identify and
incorporate all the relevant information.

The information that has, in this paper, been translated into the entropic
prior is that contained in the likelihood. The \emph{bare} entropic priors
discussed here apply to a situation where all we know about the quantities
$\theta$ is that they appear as parameters in the likelihood $p(y|\theta)$,
\emph{and nothing else}. Generalizations are, of course, possible. Sometimes
we are aware of additional relevant information beyond what is contained in
the likelihood and it can easily be incorporated into a modified entropic
prior. Other times we might be guilty of overlooking additional information we
already have. Indeed, we would not be willing to spend valuable effort in the
determination of a parameter $\theta$ unless we suspected that knowledge of
$\theta$ has important implications elsewhere. Typically we know something
about the physical significance and the physical meaning of $\theta$. It is
clear that in these cases we know considerably more than just that $\theta$ is
a parameter appearing in the likelihood. We might even conceive of several
different experiments, $e=1,2,\ldots$, each yielding different sets of data
$y_{e}$ related to $\theta$ by different likelihood functions $p_{e}%
(y_{e}|\theta)$. It is sometimes objected that one's prior knowledge about
$\theta$ should not depend on which experiment one decides to use to measure
it, but this objection is misplaced: the mere fact that $\theta$ is measurable
through one or another experiment is additional information which, if
relevant, should be taken into account.

Another family of problems that can be tackled as a rather straightforward
extension of the ideas described here involve choosing which likelihood
distribution from among several competing candidates is responsible for
generating the data. Indeed, it is clear that any systematic approach to model
selection requires as a prerequisite the capability to process in an objective
way the information implicit in each of those likelihoods. Except for some
brief remarks in the final section, all these further developments, valuable
as they might be, will be addressed elsewhere.

Our contribution includes a derivation of an entropic prior (section 3)
following the same principles of ME inference that have been successful in
statistical mechanics. In fact, our whole approach is guided by intuition
gained from applications of ME\ to statistical mechanics. Preliminary steps
along this direction were taken in \cite{Caticha00a} where a problem with the
important case of experiments that can be indefinitely repeated had already
been identified but not fully resolved. This problem, re-examined in section
4, is interpreted as a symptom that important relevant information has been
overlooked. The complete resolution, which hinges on identifying and
incorporating this additional information, is given in sections 5 and 6. The
actual way in which ME is used in the derivation, in analogy to standard
applications in statistical mechanics, turns out to be important because it
clarifies what it is that has been derived and how to use it: ours is, in
effect, a third proposal for an entropic prior. In section 7 we discuss in
detail the important example of a Gaussian likelihood and finally, in section
8, we summarize and comment on the differences among the three versions of
entropic prior and on possible further developments.

\section{The logic behind the ME method}

The goal is to update beliefs about $y\in Y$ which are codified in the prior
probability distribution $m(y)$ to a posterior distribution $p(y)$ when new
information in the form of a constraint becomes available. (The constraints
can, but need not, be linear.) The selection is carried out by ranking the
probability distributions according to increasing \emph{preference}. One
feature we impose on the ranking scheme is transitivity: if distribution
$p_{1}$ is preferred over distribution $p_{2}$, and $p_{2}$ is preferred over
$p_{3}$, then $p_{1}$ is preferred over $p_{3}$. Such transitive rankings are
implemented by assigning to each $p(x)$ a real number $S[p]$ called the
entropy of $p$ in such a way that if $p_{1}$ is preferred over $p_{2}$, then
$S[p_{1}]>S[p_{2}]$. The selected $p$ will be that which maximizes $S[p]$.
Thus the method involves entropies which are real numbers and entropies that
should be maximized. These are features imposed by design; they are dictated
by the function that the ME method is supposed to perform .

Next we determine the functional form of $S[p]$. This is the rule that defines
the ranking scheme. \emph{The purpose of the rule is to do induction.} We want
to extrapolate, to generalize from those special cases where we know what the
preferred distribution should be to the much larger number of cases where we
do not. Thus, in order to be an inductive rule $S[p]$ must have wide
applicability; we will assume that \emph{the same rule applies to all cases}.
There is no justification for this universality except for the usual pragmatic
justification of induction: we must be inclined to generalize lest we become
paralyzed into not generalizing at all. But then, we should remain cautious
and keep in mind that in many instances induction just fails.

The argument goes as follows \cite{Skilling88}. If a general theory exists,
then it must apply to special cases. Furthermore, if in a certain special case
the preferred distribution is known, then this knowledge can be used to
constrain the form of $S[p]$. Finally, if enough special cases are known, then
$S[p]$ will be completely determined. The known special cases are called the
\textquotedblleft axioms\textquotedblright\ of ME. As we will see below the
axioms reflect the conviction that one should not change one's mind
frivolously, that whatever was learned in the past is important. The chosen
posterior distribution should coincide with the prior as closely as possible
and one should only update those aspects of one's beliefs for which corrective
new evidence has been supplied. The three axioms are listed below.

\textbf{Axiom 1: Locality}. \emph{Local information has local effects.} We do
not revise the relative probabilities $p(y^{\prime})/p(y)$ with $y$ and
$y^{\prime}$ within a certain domain $D\subset Y$ unless the newly provided
information refers explicitly to the domain $D$. The power of this axiom stems
from the arbitrariness in the choice of $D$. The consequence of the axiom is
that non-overlapping domains of $y$ contribute additively to the entropy:
$S[p]=\int dy\,F(p(y))$ where $F$ is some unknown function.

\textbf{Axiom 2: Coordinate invariance.} \emph{The ranking should not depend
on the system of coordinates. }The coordinates that label the points $y$ are
arbitrary; they carry no information. The consequence of this axiom is that
$S[p]=\int dy\,p(y)f(p(y)/m(y))$ involves coordinate invariants such as
$dy\,p(y)$ and $p(y)/m(y)$, where the density $m(y)$ and the function $f$ are,
at this point, unknown.

Next we make a second use of the locality axiom to enforce objectivity. We
allow domain $D$ to extend over the whole space $Y$ and assert that \emph{when
there is no new information there is no reason to change one's mind. }When
there are no constraints the selected posterior distribution should coincide
with the prior distribution. This eliminates the arbitrariness in the density
$m(y)$: up to normalization $m(y)$ is the prior distribution.

\textbf{Axiom 3:\ Consistency for independent subsystems}. \emph{When a system
is composed of independent subsystems it should not matter whether the
inference procedure treats them separately or jointly.} If $y=(y_{1},y_{2})\in
Y=Y_{1}\times Y_{2}$, and the subsystem priors $m_{1}(y_{1})$ and $m_{2}%
(y_{2})$ are respectively upgraded to $p_{1}(y_{1})$ and $p_{2}(y_{2})$, then
the prior for the whole system $m_{1}(y_{1})m_{2}(y_{2})$ should be upgraded
to $p_{1}(y_{1})p_{2}(y_{2})$. This axiom restricts the function $f$ to be a
logarithm. (The fact that the logarithm applies also when the subsystems are
not independent follows from our inductive hypothesis that the ranking scheme
has universal applicability.)

The overall consequence of these axioms \cite{footnote2} is that probability
distributions $p(y)$ should be ranked relative to the prior $m(y)$ according
to their (relative) entropy \cite{footnote1},
\begin{equation}
S[p,m]=-\int dy\,p(y)\log\frac{p(y)}{m(y)}. \label{S[p]}%
\end{equation}
The derivation has singled out $S[p,m]$ as \emph{the unique entropy to be used
in inductive inference}. Other expressions, such as $S[m,p]$, or
$S[p,m]+S[m,p]$, or even expressions that do not involve the logarithm, may be
useful for other purposes, but they do not constitute an induction: they are
not a generalization from the simple cases described in the axioms.

We end this section with two comments on the prior density $m(y)$. First,
$S[p,m]$ may be infinitely negative when $m(y)$ vanishes within some region
$D$. In other words, the ME method confers an overwhelming preference on those
distributions $p(y)$ that vanish whenever $m(y)$ does. Is this a problem? Not
really. A similar \textquotedblleft problem\textquotedblright\ also arises in
the context of Bayes' theorem. A vanishing prior represents a tremendously
serious prejudice because no amount of data to the contrary would allow us to
revise it. The solution in both cases is to recognize that unless we are
absolutely certain that $y$ could not possibly lie within $D$ then we should
not have assigned $m(y)=0$ in the first place. Assigning a very low but non
zero prior represents a safer and less prejudiced representation of one's
beliefs and/or doubts both in the context of Bayesian and of ME inference.

Second, choosing the prior density $m(y)$ can be tricky. When there is no
information leading us to prefer one microstate of a physical system over
another we might as well assign equal prior probability to each state. Thus it
is reasonable to identify $m(y)$ with the density of states and the invariant
$m(y)dy$ is the number of microstates in $dy$. This is the basis for
statistical mechanics. Other examples of relevance to physics arise when there
is no reason to prefer one region of the space $Y$ over another. Then we
should assign the same prior probability to regions of the same
\textquotedblleft volume,\textquotedblright\ and we can choose $\int
_{R}dy\,m(y)$ to be the volume of a region $R$ in the space $Y$. Notice that
because of the presence of the prior $m(y)$ not all subjectivity has been
eliminated and Laplace's principle of insufficient reason still plays an
important role, albeit in a somewhat modified form. Just as with Bayes'
theorem, what is objective here is the manner in which information is
processed, not the initial probability assignments.

\section{Entropic priors: the basic idea}

In this section we follow \cite{Caticha00a} closely. We use the ME method to
derive a prior $\pi(\theta)$ for use in Bayes' theorem,
\begin{equation}
p(\theta|y)\propto p(y,\theta)=\pi(\theta)p(y|\theta)\,.
\end{equation}
The selection of a preferred distribution using the ME method demands that the
space in which the search will be conducted be specified. Being a consequence
of the product rule for joint probabilities, Bayes' theorem requires that
assertions such as `$y$ and $\theta$' be meaningful and that the `probability
of $y$ and $\theta$' be well defined. Therefore we must focus our attention on
$p(y,\theta)$ rather than $\pi(\theta)$; the relevant universe of discourse is
neither $\Theta$, the space of all $\theta$s, nor the data space $Y$, but the
product $\Theta\times Y$. This point, first made by Rodr\'{\i}guez
\cite{Rodriguez98}, is central to the argument. Our derivation and the final
result, however, differ from his in several respects \cite{Rodriguez98,
Rodriguez02}.

To rank distributions in the space $\Theta\times Y$ we must decide on a prior
$m(y,\theta)$. At this starting point absolutely nothing is known about the
variables $\theta$, in particular, they have no physical meaning, and no
relation between $y$ and $\theta$ is known. The $\theta$s are totally
arbitrary. Therefore the prior must be a product $m(y)\mu(\theta)$ of the
separate priors in the spaces $Y$ and $\Theta$. Indeed, the distribution that
maximizes the relative entropy
\begin{equation}
\sigma\lbrack p]=-\int dy\,d\theta\,p(y,\theta)\,\log\frac{p(y,\theta
)}{m(y)\mu(\theta)},
\end{equation}
when no constraints are imposed is $p(y,\theta)\propto m(y)\mu(\theta)$; it is
such that data about $y$ tells us absolutely nothing about $\theta$.

In what follows we assume that $m(y)$ is known. We consider this an important
part of understanding what data it is that has been collected. In section 7 we
will suggest a reasonable $m(y)$ for the special case of a Gaussian
likelihood. The prior $\mu(\theta)$ remains unspecified.

Next we incorporate the crucial piece of information from which the parameters
$\theta$ derive their physical meaning and which establishes the relation
between $\theta$ and $y$: the likelihood function $p(y|\theta)$ is known. This
has two consequences: First, the joint distribution $p(y,\theta)$ is
constrained to be of the form $\pi(\theta)p(y|\theta)$. Notice that this
constraint is not in the form that is most usual for applications of the ME
method: it is not an expectation value. Note also that the only information we
are using about the quantities $\theta$ is that they appear as parameters in
the likelihood $p(y|\theta)$, \emph{nothing else}. In many situations of
experimental interest there exists additional relevant information beyond what
is contained in the likelihood; such information should be included as
additional constraints in the maximization of the relative entropy $\sigma$.

Second, now that a bare minimum is known about $\theta$, namely that each
$\theta$ represents a probability distribution, there is a natural but still
subjective choice for $\mu(\theta)$. As discussed in \cite{Amari85}, except
for an overall multiplicative constant, there is a unique Riemannian metric
that adequately reflects the fact that the points in a space of probability
distributions are not `structureless', but happen to be probability
distributions; this is the Fisher-Rao metric. Within the finite-dimensional
subspace defined by the constraint -- the known $p(y|\theta)$ -- the natural
metric on $\Theta$ is $d\ell^{2}=g_{ij}d\theta^{i}d\theta^{j}$, where the
unique $g_{ij}$ induced by the family of distributions $p(y|\theta)$ is
\begin{equation}
g_{ij}=\int dy\,p(y|\theta)\frac{\partial\log p(y|\theta)}{\partial\theta^{i}%
}\frac{\partial\log p(y|\theta)}{\partial\theta^{j}}. \label{Fisher metric}%
\end{equation}
Accordingly we choose $\mu(\theta)=g^{1/2}(\theta)$, where $g(\theta)$ is the
determinant of $g_{ij}$. Having identified the prior measure and the
constraints, we allow the ME method to take over.

The preferred distribution $p(y,\theta)$ is chosen by varying $\pi(\theta) $
to maximize
\begin{align}
\sigma\lbrack\pi]  &  =-\int dy\,d\theta\,\pi(\theta)p(y|\theta)\,\log
\frac{\pi(\theta)p(y|\theta)}{g^{1/2}(\theta)m(y)}\label{sigma[pi]}\\
&  =-\int d\theta\,\pi(\theta)\log\frac{\pi(\theta)}{g^{1/2}(\theta)}+\int
d\theta\,\pi(\theta)S(\theta)~,\nonumber
\end{align}
where $S(\theta)$ is the entropy of the likelihood,
\begin{equation}
S(\theta)=-\int\,dy\,p(y|\theta)\log\frac{p(y|\theta)}{m(y)}. \label{Stheta}%
\end{equation}
$\,$ Writing the Lagrange multiplier that enforces $\int d\theta\,\pi
(\theta)=1$ as $1-\log\zeta$, and assuming $p(y|\theta)$ is normalized yields
\begin{equation}
0=\int\,d\theta\left(  -\log\frac{\pi(\theta)}{g^{1/2}(\theta)}+S(\theta
)-\log\zeta\right)  \,\delta\pi(\theta)\,,
\end{equation}
Therefore the probability that the value of $\theta$ should lie within the
small volume $g^{1/2}(\theta)d\theta$ is
\begin{equation}
\pi(\theta)d\theta=\frac{1}{\zeta}\,\,e^{S(\theta)}g^{1/2}(\theta)d\theta
\quad\text{with\quad}\zeta=\int d\theta\,g^{1/2}(\theta)\,e^{S(\theta)}.
\label{main}%
\end{equation}
This entropic prior is our first main result. It tells us that the preferred
value of $\theta$ is that which maximizes the entropy $S(\theta)$ because this
maximizes the scalar probability density $\exp S(\theta)$. It also tells us
the degree to which values of $\theta$ away from the maximum are ruled out; in
many cases the preference for the ME distribution can be overwhelming. Note
also that the density $\exp S(\theta)$ is a scalar function and the presence
of the Jacobian factor $g^{1/2}(\theta)$ makes eq.(\ref{main}) manifestly
invariant under changes of the coordinates $\theta$ in the space $\Theta$.

We can claim a partial success. The ingredients that have been used are
precisely those that led us to consider using Bayes' theorem in the first
place. The information contained in the model -- by which we mean that the
data space $Y$, its measure $m(y)$, and the conditional distribution
$p(y|\theta)$ -- has been translated into a prior $\pi(\theta)$. The success
is partial because it has been achieved for the special case of the fixed data
space $Y$ of those experiments which cannot conceivably be repeated. A more
complete treatment requires that we address the important case of experiments
that can be repeated indefinitely.

\section{Repeatable experiments}

Experiments need not be repeatable but sometimes they are. Let us assume that
successive repetitions are possible and that they happen to be independent.
Suppose, to be specific, that the experiment is performed twice so that the
space of data $Y\times Y=Y^{2}$ consists of the possible outcomes $y_{1}$ and
$y_{2}$. Suppose further that $\theta$ is not a \textquotedblleft
random\textquotedblright\ variable; the value of $\theta$ is fixed but
unknown. Then the joint distribution in the space $\Theta\times Y^{2}$ is
\begin{equation}
p(y_{1},y_{2},\theta)=\pi^{(2)}(\theta)\,p(y_{1},y_{2}|\theta)=\pi
^{(2)}(\theta)p(y_{1}|\theta)p(y_{2}|\theta),
\end{equation}
and the appropriate $\sigma$ entropy is
\begin{equation}
\sigma^{(2)}[\pi]=-\int dy_{1}\,dy_{2}\,d\theta\,p(y_{1},y_{2},\theta
)\,\log\frac{p(y_{1},y_{2},\theta)}{\left[  g^{(2)}(\theta)\right]
^{1/2}\,m(y_{1})m(y_{2})},
\end{equation}
where $g^{(2)}(\theta)$ is the determinant of the Fisher-Rao metric for
$p(y_{1},y_{2}|\theta)$. From Eq.(\ref{Fisher metric}) it follows that
$g_{ij}^{(2)}=2g_{ij}$ so that $g^{(2)}(\theta)=2^{d}g(\theta)$, $d$ being the
dimension of $\theta$. Maximizing $\sigma^{(2)}[\pi]$ subject to
$\int\,d\theta\,\pi^{(2)}(\theta)=1$ we get
\begin{equation}
\pi^{(2)}(\theta)=\frac{1}{Z^{(2)}}\,g^{1/2}(\theta)\,e^{S^{(2)}(\theta
)}=\frac{1}{Z^{(2)}}\,g^{1/2}(\theta)\,e^{2S(\theta)},
\end{equation}
where $S^{(2)}(\theta)=2S(\theta)$ is the entropy of $\,p(y_{1},y_{2}|\theta
)$, and $S(\theta)\overset{\operatorname*{def}}{=}S^{(1)}(\theta)$. The
generalization to $N$ repetitions of the experiment, with data space $Y^{N}$,
is immediate,
\begin{equation}
\pi^{(N)}(\theta)=\frac{1}{Z^{(N)}}\,g^{1/2}(\theta)\,e^{S^{(N)}(\theta
)}=\frac{1}{Z^{(N)}}\,g^{1/2}(\theta)\,e^{NS(\theta)}. \label{pi(n)}%
\end{equation}
This is clearly wrong: the dependence of $\pi^{(N)}$ on the amount $N$ of data
would lead us to a perpetual revision of the prior as more data is collected.
The absurdity of this situation becomes manifest when we consider the case of
large $N$. Then the exponential preference for the value of $\theta$ that
maximizes $S(\theta)$ becomes so pronounced that no amount of data to the
contrary can successfully overcome its effect. The data becomes irrelevant,
and the more data we have, the more irrelevant it becomes.

Repeatable experiments present us with a problem. One possible attitude is to
blame the ME method: it gives nonsense and cannot be trusted. As with all
inductive methods this is, of course, a logical possibility. A second, more
constructive approach, is to always be prepared to question the results of ME
calculations on the basis that there is no guarantee that all the information
relevant to the situation at hand has been taken into account. The problem is
not a failure of the ME method but a failure to include all the relevant information.

That this is indeed the case can be seen as follows: When we say an experiment
can be repeated twice, $N=2$, we actually know more than just $p(y_{1}%
,y_{2}|\theta)=\,p(y_{1}|\theta)p(y_{2}|\theta)$. We also know that forgetting
or discarding the value of say $y_{2}$, yields an experiment that is totally
indistinguishable from the single, $N=1$, experiment. This \emph{additional}
information is quantitatively expressed by $\int dy_{2}\,p(y_{1},y_{2}%
,\theta)=p(y_{1},\theta)$, or equivalently
\begin{equation}
\int dy_{2}\,\pi^{(2)}(\theta)p(y_{1}|\theta)p(y_{2}|\theta)=\pi^{(1)}%
(\theta)p(y_{1}|\theta)\,,
\end{equation}
which leads to $\pi^{(2)}(\theta)=\pi^{(1)}(\theta)$. In the general case we
get the manifestly reasonable result
\begin{equation}
\pi^{(N)}(\theta)=\pi^{(N-1)}(\theta)=\ldots=\pi^{(1)}(\theta)~.
\label{constr on pi}%
\end{equation}
The challenge then is to identify a constraint that codifies this information
within each space $\Theta\times Y^{N}$.

\section{More information: the Lagrange multiplier $\alpha$}

The problem with the prior $\pi^{(N)}(\theta)$ in eq.(\ref{pi(n)}) is that it
expresses an overwhelming preference for the value $\theta_{\max}$ of $\theta$
that maximizes the entropy $S(\theta)$. Indeed, as $N\rightarrow\infty$ we
have $\pi^{(N)}(\theta)\rightarrow\delta(\theta-\theta_{\max})$ leading to
\begin{equation}
\langle S\rangle=\int d\theta\,\pi^{(N)}(\theta)S(\theta)\overset
{N\rightarrow\infty}{\longrightarrow}S(\theta_{\max})~,
\end{equation}
which is manifestly incorrect. This suggests that a better prior would be
obtained by maximizing the entropy $\sigma^{(N)}$ of distributions on the
space space $\Theta\times Y^{N}$ subject to an additional constraint on the
numerical value $\bar{S}$ of the expected entropy $\langle S\rangle$. It is
not that we happen to know the numerical value $\bar{S}$ of $\langle S\rangle
$. In fact we do not. It is rather that we recognize that information about
$\bar{S}$ is relevant in the sense that if $\bar{S}$ were known the problem
above would not arise. Naturally, additional effort will be required to obtain
the needed value of $\bar{S}$.

The logic of the previous paragraph may sound unfamiliar and further comments
may be helpful. When justifying the use of the ME method to obtain, say, the
canonical Boltzmann-Gibbs distribution ($P_{q}\propto e^{-\beta E_{q}}$) it
has been common to say something like \textquotedblleft we seek the minimally
biased (\emph{i.e.} maximum entropy) distribution that codifies the
information we do possess (the expected energy) and nothing
else\textquotedblright. Many authors find this justification objectionable.
Indeed, they might argue, for example, that the spectrum of black body
radiation is what it is independently of whatever information happens to be
available to us. We prefer to phrase our objection differently: in most
realistic situations the expected value of the energy is not a quantity we
happen to know. Nevertheless, it is still true that maximizing entropy subject
to a constraint on this (unknown) expected energy leads to correct
predictions. Therefore, the justification behind imposing a constraint on the
expected energy cannot be that this is a quantity that happens to be known --
because it is not -- but rather that the\emph{\ }expected energy is the
quantity that \emph{should }be\emph{ }known. Even if unknown, we recognize it
as the crucial relevant information without which no successful predictions
can be made. Therefore we proceed as if this crucial information were
available and produce a formalism that contains the temperature as a free
parameter that will later have to be obtained from the experiment itself. In
other words, the temperature (or expected energy) is one additional parameter
to be inferred from the data.

The entropy on the space $\Theta\times Y^{N}$ is%

\begin{align}
\sigma^{(N)}[\pi]  &  =-\int\,d\theta\,dy^{(N)}\,\pi(\theta)p(y^{(N)}%
|\theta)\log\frac{\pi(\theta)p(y^{(N)}|\theta)}{g^{1/2}(\theta)\,m(y^{(N)}%
)}\nonumber\\
&  =-\int\,d\theta\,\pi(\theta)\log\frac{\pi(\theta)}{g^{1/2}(\theta)\,}+N\int
d\theta\,\pi(\theta)S(\theta)
\end{align}
where $S(\theta)$ given by eq.(\ref{Stheta}). (A constant factor of $N^{d/2}$
associated to the Fisher-Rao measure $g^{(N)}(\theta)$ has been omitted. It
would eventually be absorbed into the normalization of $\pi(\theta)$.) To
obtain the prior $\pi(\theta)$ we maximize $\sigma^{(N)}$ subject to
constraints on $\langle S\rangle$ and that $\pi$ be normalized,
\begin{equation}
\delta\left[  \sigma^{(N)}+(1-\log\zeta)\left(  \int d\theta\,\pi
(\theta)-1\right)  +\lambda_{N}\left(  \int d\theta\,\pi(\theta)S(\theta
)-\bar{S}\right)  \right]  =0~. \label{varying sigma N}%
\end{equation}
This gives,
\begin{equation}
\int\,d\theta\left(  -\log\frac{\pi(\theta)}{g^{1/2}(\theta)}+(N+\lambda
_{N})S(\theta)-\log\zeta\right)  \,\delta\pi(\theta)=0\,.
\end{equation}
Therefore,
\begin{equation}
\pi(\theta)=\frac{1}{\zeta}g^{1/2}(\theta)\exp\left[  (N+\lambda_{N}%
)S(\theta)\right]  ~.
\end{equation}
The undesired dependence on $N$ is eliminated if in each space $\Theta\times
Y^{N}$ the Lagrange multipliers $\lambda_{N}$ are chosen so that
$N+\lambda_{N}=\alpha$ is a constant independent of $N$. The resulting
entropic prior,
\begin{equation}
\pi(\theta|\alpha)=\frac{1}{\zeta(\alpha)}g^{1/2}(\theta)e^{\alpha S(\theta
)}~, \label{main 2}%
\end{equation}
satisfies eq.(\ref{constr on pi}). This is our second main result. The prior
$\pi(\theta|\alpha)$ codifies information contained in the likelihood
function, plus information about the expected value of the entropy of the
likelihood implicit in the hyper-parameter $\alpha$,
\begin{equation}
\bar{S}(\alpha)=\frac{d}{d\alpha}\log\zeta(\alpha)~, \label{Sbar}%
\end{equation}
with $\zeta(\alpha)$ is given by
\begin{equation}
\zeta(\alpha)=\int d\theta~g^{1/2}(\theta)e^{\alpha S(\theta)}~.
\label{zeta(alpha)}%
\end{equation}

The next and final step is figure out which $\alpha$ applies to the particular
experimental situation under consideration. The natural way to proceed is to
invoke Bayes' theorem
\begin{equation}
p(\alpha,\theta|y^{N})=\pi(\alpha)\pi(\theta|\alpha)\frac{p(y^{N}|\theta
)}{p(y^{N})}\,.
\end{equation}
The choice of a prior $\pi(\alpha)$ for $\alpha$ itself is addressed in the
next section. If we were truly interested in the actual $\alpha$, we could
marginalize over $\theta$ to obtain
\begin{equation}
p(\alpha|y^{N})=\int d\theta\,p(\alpha,\theta|y^{N})=\frac{\pi(\alpha
)}{p(y^{N})}\int d\theta\pi(\theta|\alpha)p(y^{N}|\theta)\,.
\end{equation}
But our interest in the value of $\alpha$ is only indirect; $\alpha$ is a
necessary but annoying technical complication along the way to the real goal
which is inferring $\theta$. Marginalizing over $\alpha$, we get
\begin{equation}
p(\theta|y^{N})=\int d\alpha\,p(\alpha,\theta|y^{N})=\bar{\pi}(\theta
)\frac{p(y^{N}|\theta)}{p(y^{N})} \label{entropic Bayes}%
\end{equation}
where
\begin{equation}
\bar{\pi}(\theta)=\int d\alpha\,\pi(\alpha)\pi(\theta|\alpha). \label{main 3}%
\end{equation}
This is the answer we sought: the effective prior for $\theta$, the averaged
$\bar{\pi}(\theta)$, is independent of the actual data $y^{N}$, as it should.
The last step is the assignment of $\pi(\alpha)$.

\section{An entropic prior for $\alpha$}

To remain consistent with the spirit of this paper, namely using ME to obtain
priors, the prior for $\alpha$ must itself be an entropic prior. The
motivation behind discussing entropic priors is that we wish to consider
information included in the likelihood function. Since $p(y|\theta)$ refers to
$\theta$ but makes no reference to any hyper-parameters it is quite clear that
$\alpha$ should not be treated like the other $\theta$s. The relation between
$\alpha$ and the data $y$ is indirect: $\alpha$ is related to $\theta$, and
$\theta$ is related to $y$. Once $\theta$ is given, the data $y$ becomes
irrelevant, it contains no further information about $\alpha$. The whole
significance of $\alpha$ is derived purely from its appearance in $\pi
(\theta|\alpha)$, eq.(\ref{main 2}). Therefore, the relevant universe of
discourse is $A\times\Theta$ with $\alpha\in A$. We focus our attention on the
joint distribution
\begin{equation}
\pi(\alpha,\theta)=\pi(\alpha)\pi(\theta|\alpha)\,. \label{constr ATh}%
\end{equation}
and we obtain $\pi(\alpha)$ by maximizing the entropy
\begin{equation}
\Sigma\lbrack\pi]=-\int d\alpha\,d\theta\,\,\pi(\alpha,\theta)\log\frac
{\pi(\alpha,\theta)}{\gamma^{1/2}(\alpha)\,g^{1/2}(\theta)}\,
\label{Sigma prime}%
\end{equation}
where $\gamma^{1/2}(\alpha)$ is determined below. Since no reference is made
to repeatable experiments in $Y^{N}$ there is no need for any further
constraints except for normalization.

The Fisher-Rao measure $\gamma^{1/2}(\alpha)$ in eq.(\ref{Sigma prime}) is
\begin{equation}
\gamma(\alpha)=\int d\theta\,\pi(\theta|\alpha)\left[  \frac{d}{d\alpha}%
\log\pi(\theta|\alpha)\right]  ^{2}.
\end{equation}
Using eqs.(\ref{main 2}),(\ref{Sbar}) and (\ref{zeta(alpha)}) we get
\begin{equation}
\gamma(\alpha)=\int d\theta\,\pi(\theta|\alpha)\left[  S(\theta)-\frac
{d\log\zeta(\alpha)}{d\alpha}\right]  ^{2}=(\Delta S)^{2},
\end{equation}
but
\begin{equation}
\frac{d\bar{S}(\alpha)}{d\alpha}=\frac{d}{d\alpha}\frac{1}{\zeta(\alpha)}%
\frac{d\zeta(\alpha)}{d\alpha}=\frac{1}{\zeta(\alpha)}\frac{d^{2}\zeta
(\alpha)}{d\alpha^{2}}-\left[  \frac{1}{\zeta(\alpha)}\frac{d\zeta(\alpha
)}{d\alpha}\right]  ^{2}=(\Delta S)^{2}.
\end{equation}
Therefore,
\begin{equation}
\gamma(\alpha)=\frac{d^{2}\log\zeta(\alpha)}{d\alpha^{2}}\,.
\label{gamma(alpha)}%
\end{equation}
The interpretation is straightforward: the distance between $\pi(\theta
|\alpha)$ and $\pi(\theta|\alpha+d\alpha)$ is given by
\begin{equation}
\gamma^{1/2}(\alpha)d\alpha=\Delta S\,(\alpha)d\alpha\,,
\end{equation}
or, in words, the local entropy uncertainty $\Delta S$ is the distance per
unit change in $\alpha$.

To maximize $\Sigma$ rewrite it as
\begin{equation}
\Sigma\lbrack\pi]=-\int d\alpha\,\pi(\alpha)\log\frac{\pi(\alpha)}%
{\gamma^{1/2}}+\int d\alpha\,\pi(\alpha)\,s(\alpha),
\end{equation}
where $s(\alpha)$ is given by
\begin{align}
s(\alpha)  &  =-\int d\theta\,\pi(\theta|\alpha)\log\frac{\pi(\theta|\alpha
)}{g^{1/2}(\theta)}\,\label{s(alpha)}\\
&  =\log\zeta(\alpha)-\alpha\frac{d\log\zeta(\alpha)}{d\alpha}~.\nonumber
\end{align}
Then, varying with respect to $\pi(\alpha)$ gives
\begin{equation}
\pi(\alpha)=\frac{1}{z}\gamma^{1/2}(\alpha)e^{s(\alpha)}\,. \label{main 4}%
\end{equation}
This is our third main result. It completes our derivation of the actual prior
for $\theta$: the averaged $\bar{\pi}(\theta)$ in eq.(\ref{main 3}) codifies
information contained in the likelihood function, plus the insight that for
repeatable experiments, information about the expected likelihood entropy,
even if unavailable, is relevant.

We argued above that the hyper-parameter $\alpha$ should not be treated in the
same way as the other parameters $\theta$ because the likelihood $\pi
(y|\theta)$ refers only to $\theta$s and not to $\alpha$. Nonetheless, it may
still be worthwhile to discuss briefly what would happen if $\alpha$\ were
treated as one of the $\theta$s. In this case, the entropic prior $\pi
(\alpha)$ would be determined by focusing our attention on the joint
distribution
\begin{equation}
p(\alpha,\theta,y^{N})=\pi(\alpha)\pi(\theta|\alpha)p(y^{N}|\theta)\,,
\end{equation}
where the last two factors on the right are assumed known. The assumed
universe of discourse would be $A\times\Theta\times Y^{N}$. A straightforward
application of the ME method would, as before, run into trouble with an
unwanted $N$ dependence which would require the introduction of a new
constraint on the appropriate expected entropy. Thus, the entropic prior for
$\alpha$ would involve a second hyper-parameter $\alpha_{2}$. The unknown
$\alpha_{2}$ would itself require its own entropic prior, involving yet a
third hyper-parameter $\alpha_{3}$, and so on. There would be an endless chain
of hyper-parameters \cite{Rodriguez89}. In any practical calculation, the
chain would have to be truncated. Whether the predictions about $\theta$
depend on where and how the truncation is carried out remains to be studied.
But, fortunately, this is not necessary: $\alpha$ is not like the other
$\theta$s.

\section{Example: a Gaussian model}

Consider data $y^{N}=\{y_{1},\ldots,y_{N}\}$ that are scattered around an
unknown value $\mu$,
\begin{equation}
y=\mu+\nu~
\end{equation}
with $\langle\nu\rangle=0$ and $\langle\nu^{2}\rangle=\sigma^{2}.$ The goal is
to estimate the parameters $\theta=(\theta^{1},\theta^{2})=(\mu,\sigma)$ on
the basis of the data $y^{N}$ and the information implicit in the model: the
data space $Y$, the measure $m(y)$ (discussed below), and the Gaussian
likelihood,
\begin{equation}
p(y|\mu,\sigma)=\frac{1}{\left(  2\pi\sigma^{2}\right)  ^{1/2}}\exp\left[
-\frac{(y-\mu)^{2}}{2\sigma^{2}}\right]  ~. \label{Gaussian likelihood}%
\end{equation}

In section 3 we asserted that knowing the measure $m(y)$ is part of knowing
what data has been collected. Therefore, nothing can be said about $m(y)$
without further specification of the experimental situation. It turns out,
however, that in many physical situations where the data happen to be
distributed according to eq.(\ref{Gaussian likelihood}) the underlying space
$Y$ is sufficiently symmetric, \emph{i.e.}, invariant under translations, that
we can assume $m(y)=m=\operatorname{constant}.$ This is physically reasonable.
Gaussian distributions arise when the measured value of $y$ is the sum of a
large number of \textquotedblleft microscopic\textquotedblright\ contributions
and the details of how the individual contributions are themselves distributed
are washed out in the \textquotedblleft macroscopic\textquotedblright\ sum.
The macroscopically relevant features are just those that distinguish one
Gaussian from another, namely, the mean $\mu$ and the variance $\sigma^{2}$.
This is the physical basis behind the Central Limit Theorem. But if
microscopic details are irrelevant it should be possible to understand the
situation from a purely macroscopic point of view: it should be possible to
obtain the Gaussian distribution as the preferred one among all those with the
given $\mu$ and $\sigma^{2}$, and this is, indeed, the case: setting
$m(y)=\operatorname{constant}$ in $S[p,m]$, eq.(\ref{S[p]}), and maximizing
subject to constraints on the mean and variance yields
eq.(\ref{Gaussian likelihood}).

From eqs. (\ref{Stheta}) and (\ref{Gaussian likelihood}) the entropy of the
likelihood is
\begin{equation}
S(\mu,\sigma)=\log\left[  \frac{\sigma}{\sigma_{0}}\right]  \quad
\text{where}\quad\sigma_{0}\overset{\operatorname*{def}}{=}\left(  \frac
{e}{2\pi}\right)  ^{1/2}\frac{1}{m}~,
\end{equation}
and the corresponding Fisher-Rao measure, from eq.(\ref{Fisher metric}) is
\begin{equation}
g(\mu,\sigma)=\det\left\vert
\begin{array}
[c]{cc}%
1/\sigma^{2} & 0\\
0 & 2/\sigma^{2}%
\end{array}
\right\vert =\frac{2}{\sigma^{4}}~.
\end{equation}

Note that both $S(\mu,\sigma)$ and $g(\mu,\sigma)$ are independent of $\mu$.
This means that if we were concerned with the simpler problem of estimating
$\mu$\ in a situation where $\sigma$ happens to be known, then the entropic
prior, in any of the versions eq.(\ref{main}), (\ref{main 2}), or
(\ref{main 3}), is a constant independent of $\mu$. In other words, when
$\sigma$ is known, the Bayesian estimate of $\mu$ using entropic priors
coincides with the maximum likelihood estimate, \emph{i.e.}, by the popular
procedure of minimizing
\begin{equation}
\chi^{2}=\frac{1}{\mu}\sum_{i=1}^{N}\left(  y_{i}-\mu\right)  ^{2}~.
\end{equation}

Returning to the more interesting case of unknown $\sigma$, the $\alpha
$-dependent entropic prior, eq.(\ref{main 2}) is
\begin{equation}
\pi(\mu,\sigma|\alpha)=\frac{2^{1/2}}{\zeta(\alpha)}~\frac{\sigma^{\alpha-2}%
}{\sigma_{0}^{\alpha}}~. \label{pi(theta|alpha)G0}%
\end{equation}
$\pi(\mu,\sigma|\alpha)$ is improper in both $\mu$ and $\sigma$; normalization
requires\ the introduction of high and low cutoffs for both $\mu$ and $\sigma
$. The fact that without cutoffs the model is not well defined is an
indication that more relevant information is being requested: the cutoffs
constitute relevant information that must be taken into account. (The logic
parallels that which led to the introduction of $\alpha$ in section 5.)\ The
case of unknown cutoff values is important and we intend to explore it in
detail in future work. The basic idea is that specifying cutoffs is an
integral part of defining the model, and therefore the choice of cutoffs can
be tackled as a problem of model selection. In the remainder of this section,
however, we will assume that the information about cutoffs is already available.

It is convenient to write the range of $\mu$ as $\Delta\mu=\mu_{H}-\mu_{L}$
and to define the $\sigma$ cutoffs in terms of dimensionless quantities
$\varepsilon_{L}$ and $\varepsilon_{H}$; $\sigma$ extends from $\sigma
_{L}=\sigma_{0}\varepsilon_{L}$ to $\sigma_{H}=\sigma_{0}/\varepsilon_{H}$.
Then $\zeta(\alpha)$ and $\pi(\mu,\sigma|\alpha)$ are given by
\begin{equation}
\zeta(\alpha)=\frac{2^{1/2}\Delta\mu}{\sigma_{0}}\frac{\varepsilon
_{H}^{1-\alpha}-\varepsilon_{L}^{\alpha-1}}{\alpha-1}~.
\end{equation}
and
\begin{equation}
\pi(\mu,\sigma|\alpha)=\frac{1}{\Delta\mu\sigma_{0}}\frac{\alpha
-1}{\varepsilon_{H}^{1-\alpha}-\varepsilon_{L}^{\alpha-1}}~\left(
\frac{\sigma}{\sigma_{0}}\right)  ^{\alpha-2}~.
\label{pi(theta|alpha)Gaussian}%
\end{equation}
Notice that in the special case of $\alpha=1$, the prior over $\sigma$ reduces
to $d\sigma/\sigma$ which is called the Jeffreys prior and is usually
introduced by the requirement of invariance under scale transformations,
$\sigma\rightarrow\lambda\sigma$.

Writing $\varepsilon\overset{\operatorname*{def}}{=}(\varepsilon
_{L}\varepsilon_{H})^{1/2}$, the prior for $\alpha$ can be obtained from
eq.(\ref{gamma(alpha)}),
\begin{equation}
\gamma(\alpha)=\frac{1}{(\alpha-1)^{2}}-\left(  \frac{2\log\varepsilon
}{\varepsilon^{1-\alpha}-\varepsilon^{\alpha-1}}\right)  ^{2}
\label{gammaGaussian}%
\end{equation}
and from eqs.(\ref{main 3}) and (\ref{s(alpha)}),
\begin{equation}
\pi(\alpha)=\frac{\gamma^{1/2}(\alpha)}{z}\frac{\varepsilon^{1-\alpha
}-\varepsilon^{\alpha-1}}{\alpha-1}\exp\left[  \frac{1}{\alpha-1}+\alpha
\frac{\varepsilon^{1-\alpha}+\varepsilon^{\alpha-1}}{\varepsilon^{1-\alpha
}-\varepsilon^{\alpha-1}}\log\varepsilon\right]  ~, \label{pi(alpha)Gaussian}%
\end{equation}
where the normalization $z$ has been suitably redefined.

\begin{figure}
[ptb]
\begin{center}
\includegraphics[
height=2.5554in,
width=3.2446in
]%
{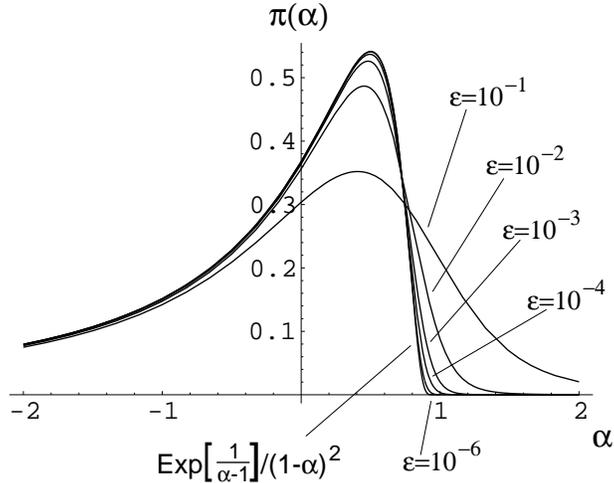}%
\caption{The prior $\pi(\alpha)$ for various values of the cutoff parameter
$\varepsilon$, as $\varepsilon\rightarrow0$. }%
\label{Fig. 1}%
\end{center}
\end{figure}

Eqs.(\ref{gammaGaussian}) and (\ref{pi(alpha)Gaussian}) simplify considerably
when we take the limit $\varepsilon\rightarrow0$. Clearly the same result is
obtained whether we let $\varepsilon_{H}\rightarrow0$ while keeping
$\varepsilon_{L}$ fixed, or letting $\varepsilon_{L}\rightarrow0$ while
keeping $\varepsilon_{H}$ fixed, or even allowing $\varepsilon_{H}%
\rightarrow0$ and $\varepsilon_{L}\rightarrow0$ simultaneously. The resulting
$\gamma(\alpha)$ and $\pi(\alpha)$ are%
\begin{equation}
\gamma(\alpha)=\frac{1}{(\alpha-1)^{2}}~,
\end{equation}
and
\begin{equation}
\pi(\alpha)=\left\{
\begin{array}
[c]{cc}%
\frac{1}{\left(  1-\alpha\right)  ^{2}}\exp\left[  {\frac{1}{\alpha-1}}\right]
& \text{for\quad}\alpha<1\\
0 & \text{for\quad}\alpha\geq1
\end{array}
\right.  \label{limit piGaussian}%
\end{equation}
where $\pi(\alpha)$ is normalized. This is shown in Fig. \ref{Fig. 1}.%

$\pi(\alpha)$ reaches its maximum value at $\alpha=1/2$. Since $\pi
(\alpha)\sim\alpha^{-2}$ for $\alpha\rightarrow-\infty$ the expected value of
$\alpha$ and all higher moments diverge. This suggests that replacing the
unknown $\alpha$ in the prior $\pi(\theta|\alpha)$ by any given numerical
value $\hat{\alpha}$ is probably not a good approximation.\ 

As explained in section 5, since $\alpha$ is unknown, the effective prior for
$\theta=(\mu,\sigma)$ is obtained marginalizing $\pi(\mu,\sigma,\alpha
)=\pi(\mu,\sigma|\alpha)\pi(\alpha)$ over $\alpha$, eq.(\ref{main 3}). Since
$\pi(\alpha)=0$ for $\alpha\geq1$ as $\varepsilon\rightarrow0$ we can safely
take the limit $\varepsilon_{H}\rightarrow0$ or $\sigma_{H}\rightarrow\infty$.
Conversely, since $\pi(\alpha)\neq0$ for $\alpha<1$ we cannot take
$\varepsilon_{L}\rightarrow0$ or $\sigma_{L}\rightarrow0$. The limit
$\sigma_{H}\rightarrow\infty$ while keeping $\sigma_{L}$ fixed gives,
\begin{equation}
\pi(\mu,\sigma,\alpha)=\left\{
\begin{array}
[c]{cc}%
\frac{1}{\Delta\mu\sigma_{L}}\frac{\exp\left[  {\frac{1}{\alpha-1}} \right]  }
{1-\alpha}\left(  \frac{\sigma}{\sigma_{L}}\right)  ^{\alpha-2} &
\text{for\quad}\alpha<1\\
0 & \text{for\quad}\alpha\geq1.
\end{array}
\right.  \label{pi(theta,alpha)G}%
\end{equation}

\begin{figure}
[tb]
\begin{center}
\includegraphics[
height=2.6113in,
width=3.3829in
]%
{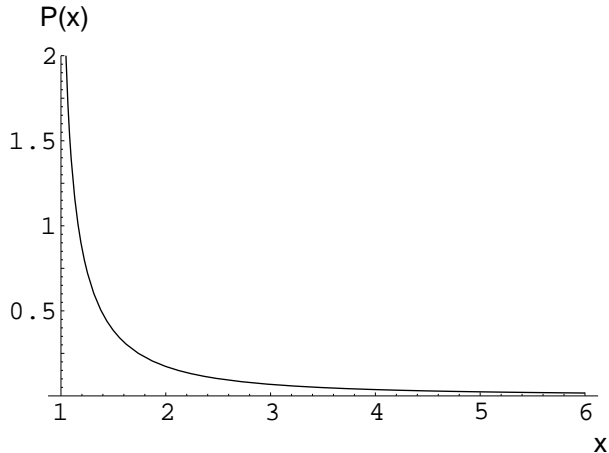}%
\caption{The effective $\bar{\pi}(\mu,\sigma)$ is shown as $P(x)=\frac{2}%
{x}K_{0}\left(  2\sqrt{\log(x)}\right)  $ where $x=\sigma/\sigma_{L}$. }%
\label{Fig. 2}%
\end{center}
\end{figure}

The averaged prior for $\mu$ and $\sigma$ is
\begin{equation}
\bar{\pi}(\mu,\sigma)=\frac{1}{\Delta\mu\sigma_{L}}\left(  \frac{\sigma_{L}%
}{\sigma}\right)  ^{2}\int_{-\infty}^{1}\frac{\exp\left[  {\frac{1}{\alpha-1}%
}\right]  }{1-\alpha}\left(  \frac{\sigma}{\sigma_{L}}\right)  ^{\alpha
}d\alpha~,
\end{equation}
which integrates to
\begin{equation}
\bar{\pi}(\mu,\sigma)=\frac{2}{\Delta\mu\sigma}~K_{0}\left(  2\sqrt{\log
\frac{\sigma}{\sigma_{L}}}\right)  ~, \label{main 5}%
\end{equation}
where $K_{0}$ is a modified Bessel function of the second kind. This is the
entropic prior for the Gaussian model. The function
\begin{equation}
P(x)=\frac{2}{x}K_{0}\left(  2\sqrt{\log x}\right)
\end{equation}
is shown in Fig. \ref{Fig. 2} as a function of $x=\sigma/\sigma_{L}$.%

$P(x)$ has an integrable singularity as $x\rightarrow1$ where it behaves as
\begin{equation}
P(x)\approx\frac{2}{x}\left(  -\log\sqrt{\log x}-\gamma\right)  \quad
\text{for}\quad x\approx1~.
\end{equation}
Since $\sigma_{L}$ is a lower cutoff the region of large $x$ is more relevant.
The leading asymptotic behavior is given by%
\begin{equation}
P(x)\approx\frac{\sqrt{\pi}}{x\left(  \log x\right)  ^{1/4}}\exp\left(
-2\sqrt{\log x}\right)  \quad\text{for}\quad x\gg1.
\end{equation}

Finally, we turn to Bayes' theorem, eq.(\ref{entropic Bayes}), with the prior
(\ref{main 5}) to obtain estimators for $\mu$ and $\sigma$. For large $N$ the
results are independent of the prior and the estimators coincide with the
standard maximum likelihood results. The case when $N$ is not so large is the
more interesting one. As estimators we can take the expected values
$\langle\mu\rangle$ and $\langle\sigma^{2}\rangle$ over the posterior
(\ref{entropic Bayes}). The integrations can be performed numerically and are
not particularly illuminating. Alternatively, one can follow standard practice
and marginalize eq.(\ref{entropic Bayes}) over $\sigma$ to obtain the
distribution $p(\mu|y^{N})$ and calculate the estimator $\hat{\mu}$ from
\begin{equation}
\left.  \frac{d}{d\mu}\log p(\mu|y^{N})\right\vert _{\hat{\mu}}=0~,
\end{equation}
and its error bar $\hat{\sigma}$ from
\begin{equation}
\left.  -\frac{d^{2}}{d\mu^{2}}\log p(\mu|y^{N})\right\vert _{\hat{\mu}}%
=\frac{1}{\hat{\sigma}^{2}}~.
\end{equation}
When $p(\mu|y^{N})$ happens to be a Gaussian these estimators coincide with
the expected values $\langle\mu\rangle$ and $\langle\sigma^{2}\rangle$. The
final result for $\hat{\mu}$ is very simple. For any value of $N$ we have
\begin{equation}
\hat{\mu}=\frac{1}{N}\sum_{i}y_{i}=\bar{y}~,
\end{equation}
the estimator $\hat{\mu}$ is the sample average. The result for $\hat{\sigma}$
is not as elegant but, of course, for large $N$ it asymptotically reduces to
$\hat{\sigma}^{2}\approx(\overline{y^{2}}-\bar{y}^{2})/N.$

\section{Final remarks}

In this paper the method of maximum relative entropy has been used to
translate the information contained in the known form of the likelihood into a
prior distribution for Bayesian inference. The argument follows closely the
analogous ME\ methods that have been so successful in statistical mechanics.
For experiments that cannot be repeated the resulting \textquotedblleft
entropic prior\textquotedblright\ is formally identical with the Einstein
fluctuation formula. For repeatable experiments, however, the expected value
of the entropy of the likelihood -- represented in terms of a Lagrange
multiplier $\alpha$ -- turns out to be relevant information that must be
included in the analysis. As an illustration the important case of a Gaussian
likelihood was treated in detail.

It may be useful to comment briefly on the differences between our entropic
prior and the versions previously proposed by Skilling and by Rodr\'{\i}guez.
Perhaps the main difference with Skilling's prior is that, unlike ours, its
use is not restricted to probability distributions but is intended for generic
\textquotedblleft positive additive distributions\textquotedblright%
\ including, for example, the distributions of intensities in images
\cite{Skilling89}. One problem here is that of justifying the applicability of
the ME method in such a general context. Our impulse to generalize is a
dangerous one; we may get away with indulging it occasionally but
overindulgence will certainly lead to error. In any case, our argument in
section 3, which consists in maximizing the entropy $\sigma$ subject to a
constraint $p(y,\theta)=\pi(\theta)p(y|\theta)$, makes no sense in the case of
generic positive additive distributions for which there is no available
product rule. A more specific problem arises from the fact that Skilling's
entropy is not, in general, dimensionless and the hyper-parameter $\alpha$ is
vaguely interpreted some sort of cutoff carrying the appropriate corrective
units. Some of the difficulties, which led Skilling to seek an alternative
approach, were identified in \cite{Skilling96}.

Rodr\'{\i}guez's approach is closer to ours. His prior applies to probability
distributions and appears to be derived from a ME principle \cite{Rodriguez02}%
. One difference, perhaps a minor one, is his treatment of the underlying
measure $m(y)$. For us $m(y)$ is not arbitrary; knowing $m(y)$ is part of
knowing what data has been collected. For him $m(y)$ is just an initial guess
and he suggests setting $m(y)=p(y|\theta_{0})$ for some value $\theta_{0}$.
The more important difference, however, is that the number of observed data
$n$ is deliberately and explicitly left unspecified. The space $\Theta\times
Y^{n}$ over which distributions are defined, and therefore the distributions
themselves, also remain unspecified. It is not clear what the maximization of
an entropy over such unspecified spaces could possibly mean but a
hyper-parameter $\alpha$ is eventually introduced and it is interpreted as a
\textquotedblleft virtual number of observations supporting the initial guess
$\theta_{0}$.\textquotedblright\ He proposes that $\alpha$ be considered as
one more among the parameters $\theta$ to be inferred. As mentioned earlier
this leads to the introduction of an endless chain of additional hyper-parameters.

There are several directions in which the ideas of this paper can be further
extended. First, we emphasize once again that the entropic priors discussed
here apply to a situation where all we know about the quantities $\theta$ is
that they appear as parameters in the likelihood $p(y|\theta)$, \emph{and
nothing else}. In many situations of experimental interest there exists
additional relevant information beyond what is contained in the likelihood.
Such information should be included as additional constraints in the
maximization of the relative entropy $\sigma$ in eq.(\ref{varying sigma N}).
The resulting modified entropic prior would provide a better representation of
our state of knowledge prior to the acquisition of the data. Indeed, the
advantage of the Bayesian approach over the usual method of maximum likelihood
is the possibility of including additional relevant information by replacing a
flat prior by an appropriately more informative prior. There is nothing to
prevent us from performing a similar improvement and going beyond the
\textquotedblleft bare\textquotedblright\ entropic priors discussed in this
paper. Two kinds of additional information that are easy to include are
restrictions on the range of the parameters $\theta$ and information about the
known expected values of some variables $a(\theta)$. Steps in this direction
were taken in section 5, where $a(\theta)$ is the likelihood entropy, and in
section 7 where high and low cutoffs on the range of the Gaussian parameters
were introduced.

Second, in the introduction we mentioned the interesting possibility of
analyzing data $y_{e}$ from different experiments, $e=1,2,\ldots$, related to
$\theta$ by different likelihood functions $p_{e}(y_{e}|\theta)$. Clearly this
can be analyzed as a single combined experiment with likelihood $p(y_{1}%
,y_{2},\ldots|\theta)=p_{1}(y_{1}|\theta)p_{2}(y_{2}|\theta)\ldots$ to which
all our previous results apply. As we stated earlier, the mere fact that
$\theta$ is measurable through one or another experiment is additional
relevant information that can be taken into account.

Third, we also mentioned that problems of model selection can be tackled as an
extension of the ideas described in this paper. On the basis of data $y$ we
want to select one model among several competing candidates labeled by
$m=1,2,\ldots$ with likelihood distributions given by $p(y|m,\theta_{m})$. The
answer, \emph{i.e.}, the probability of model $m$ given the data $y$, is given
by Bayes' theorem,
\begin{align}
p(m|y)  &  =\frac{\pi(m)}{p(y)}p(y|m)=\frac{\pi(m)}{p(y)}\int d\theta
_{m}~p(y,\theta_{m}|m)\nonumber\\
&  =\frac{1}{p(y)}\int d\theta_{m}~\pi(m,\theta_{m})~p(y|m,\theta_{m})~.
\end{align}
This is exact. The problem is solved, at least in principle, once an entropic
prior for $\pi(m,\theta_{m})$ is assigned. However, the remaining practical
problems associated with carrying out the actual numerical calculations could,
of course, still be quite formidable.

Finally, we end with a word of caution. As in all instances of inductive
inference there is the possibility that predictions based on the ME method
could be wrong because not all the information relevant to the problem at hand
was taken into account. This potential problem is not peculiar to the ME
method, it is a problem shared by all methods of induction. Nevertheless, we
are confident that the rewards of extending the benefits of an inductive
method singled out by requirements of objectivity, the ME method, beyond its
traditional territory of statistical mechanics and into that of data analysis
will be enormous.

\noindent\textbf{Acknowledgments- }Many of our comments and arguments have
been inspired by Carlos C. Rodr\'{\i}guez, Volker Dose, and Rainer Fisher
through insightful questions and discussions which we gratefully acknowledge.
A. C. also acknowledges the hospitality of the Max-Planck-Institut f\"{u}r
Plasmaphysik during the two extended visits when most of this work was carried out.


\begin{thebibliography}{99}                                                                                               %


\bibitem {Dose03}For a recent review see V.\emph{ }Dose, \textquotedblleft
Bayesian inference in physics: case studies\textquotedblright,
Rep.\ Prog.\ Phys., accepted for publication (2003); for a pedagogical
introduction see D. S. Sivia, \textquotedblleft Data Analysis, A Bayesian
Tutorial\textquotedblright\ (Oxford University Press, Oxford, 1996).

\bibitem {Feigelson92}``Statistical Challenges in Modern Astronomy I--III'',
series ed.\ by E.\ D.\ Feigelson and G.\ J.\ Babu, Springer, New York (1992,
1997, 2002).

\bibitem {Dose98}V.\ Dose, R.\ Preuss and W.\ von der Linden,
Phys.\ Rev.\ Lett.\ \textbf{81}, 3407 (1998).

\bibitem {DAgostini99}``Bayesian Reasoning in high energy physics: principles
and applications'', G.\ D'Agostini, CERN Yellow Report 99-03 (1999).

\bibitem {vdLinden93}W.\ von der Linden, M.\ Donath, and V.\ Dose,
Phys.\ Rev.\ Lett.\ \textbf{71}, 899 (1993).

\bibitem {Preuss94}R.\ Preuss et al., Phys.\ Rev.\ Lett.\ \textbf{73}, 732
(1994); R.\ Preuss, W.\ Hanke and W.\ von der Linden,
Phys.\ Rev.\ Lett.\ \textbf{75}, 1344 (1995); R.\ Preuss, W.\ Hanke,
C.\ Gr\"{o}ber, and H.G.\ Evertz, Phys.\ Rev.\ Lett.\ \textbf{79}, 1122 (1997).

\bibitem {Fischer97}R.\ Fischer, M.\ Mayer, W.\ von der Linden, and V.\ Dose,
Phys.\ Rev.\ E \textbf{56}, 6667 (1997).

\bibitem {vdLinden99}W.\ von der Linden, V.\ Dose, J.\ Padayachee, and
V.\ Prozesky, Phys.\ Rev.\ E \textbf{59}, 6527 (1999); R.\ Fischer,
K.\ M.\ Hanson, V.\ Dose, and W.\ von der Linden. Phys.\ Rev.\ E \textbf{61},
1152 (2000).

\bibitem {Schwarz01}T.\ Schwarz-Selinger, R.\ Preuss, V.\ Dose and W.\ von der
Linden, J.\ Mass Spect.\ \textbf{36}, 866 (2001).

\bibitem {vToussaint99}U.\ V.\ Toussaint, R.\ Fischer, K.\ Krieger, and
V.\ Dose. New J.\ Phys.\ \textbf{1}, 11 (1999).

\bibitem {Bretthorst88}G. L. Bretthorst, \emph{Bayesian Spectrum Analysis and
Parameter Estimation} (Lect. Notes in Physics, Vol. \textbf{48},
Springer-Verlag, Berlin, 1988); J.\ Magn.\ Reson.\ \textbf{88}, 552 (1990).

\bibitem {Kass96}For a review with annotated bibliography see \emph{e.g.}, R.
E. Kass and L. Wasserman, J. Am. Stat. Assoc. \textbf{91}, 1343 (1996).

\bibitem {Bernardo97}J. M. Bernardo, T. Z. Irony, N. D. Singpurwalla, J. Stat.
Plan. Inf. \textbf{65}, 159 (1997).

\bibitem {Jaynes68}E. T. Jaynes, IEEE Trans. Syst. Sci. Cybern. Vol.
\textbf{SSC-4}, 227 (1968); J. M. Bernardo, J. Roy. Stat. Soc. B \textbf{41},
113 (1979); A. Zellner, ``Bayesian methods and entropy in economics and
econometrics'' in \emph{Maximum Entropy and Bayesian Methods}, edited by W. T.
Grandy Jr. and L. H. Schick (Kluwer, Dordrecht, 1991).

\bibitem {Skilling89}J. Skilling, ``Classic Maximum Entropy'' in \emph{Maximum
Entropy and Bayesian Methods}, J. Skilling (ed.) (Kluwer, Dordrecht, 1989);
``Quantified Maximum Entropy'' in \emph{Maximum Entropy and Bayesian Methods},
P. F. Foug\`{e}re (ed.) (Kluwer, Dordrecht, 1990).

\bibitem {Rodriguez89}C. C. Rodr\'{\i}guez, \textquotedblleft The metrics
generated by the Kullback number\textquotedblright\ in \emph{Maximum Entropy
and Bayesian Methods}, J. Skilling (ed.) (Kluwer, Dordrecht, 1989);
\textquotedblleft Objective Bayesianism and geometry\textquotedblright\ in
\emph{Maximum Entropy and Bayesian Methods}, P. F. Foug\`{e}re (ed.) (Kluwer,
Dordrecht, 1990); \textquotedblleft Entropic priors\textquotedblright\ in
\emph{Maximum Entropy and Bayesian Methods}, edited by W. T. Grandy Jr. and L.
H. Schick (Kluwer, Dordrecht, 1991); \textquotedblleft Bayesian robustness: a
new look from geometry\textquotedblright\ in \emph{Maximum Entropy and
Bayesian Methods}, G. R. Heidbreder (ed.) (Kluwer, Dordrecht, 1996).

\bibitem {footnote1}On terminology: The terms `prior' and `posterior' are
normally used in the context of Bayes' theorem; we retain the same terminology
when using ME because we are concerned with the similar goal of processing
information to update from a prior to a posterior. The \textquotedblleft
method of ME\textquotedblright\ is usually understood in the restricted sense
that one updates from a prior distribution that happens to be uniform. Here we
adopt a broader meaning that includes updates from arbitrary priors and which
involves the maximization of relative entropy. Indeed since all entropies are
relative to some prior the qualifier `relative' is not needed and will
henceforth be omitted.

\bibitem {Shannon48}C. E. Shannon, Bell Systems Tech. Journal \textbf{27},
379, 623 (1948); C. E. Shannon and W. Weaver, \emph{The Mathematical Theory of
Communication} (Univ. of Illinois Press, Urbana, 1949); N. Wiener,
\emph{Cybernetics} (MIT Press, Cambridge, 1948); L. Brillouin, \emph{Science
and Information Theory}, (Academic Press, New York, 1956); S. Kullback,
\emph{Information Theory and Statistics} (Wiley, New York, 1959).

\bibitem {Jaynes57}E. T. Jaynes, \textquotedblleft Information Theory and
Statistical Mechanics\textquotedblright\ Phys. Rev. \textbf{106}, 620 and
\textbf{108}, 171 (1957); R. D. Rosenkrantz (ed.), \emph{E. T. Jaynes: Papers
on Probability, Statistics and Statistical Physics} (Reidel, Dordrecht, 1983);
E. T. Jaynes, \emph{Probability Theory: The Logic of Science} (Cambridge
University Press, Cambridge, 2003).

\bibitem {ShoreJohnson80}J. E. Shore and R. W. Johnson, \textquotedblleft
Axiomatic derivation of the Principle of Maximum Entropy and the Principle of
Minimum Cross-Entropy,\textquotedblright\ IEEE Trans. Inf. Theory
\textbf{IT-26}, 26 (1980); Y. Tikochinsky, N. Z. Tishby and R. D. Levine,
Phys. Rev. Lett. \textbf{52}, 1357 (1984) and Phys. Rev. \textbf{A30}, 2638
(1984); I. Csiszar, Ann. Stat. \textbf{19}, 2032 (1991).

\bibitem {Skilling88}J. Skilling, ``The Axioms of Maximum Entropy'' in
\emph{Maximum-Entropy and Bayesian Methods in Science and Engineering}, G. J.
Erickson and C. R. Smith (eds.) (Kluwer, Dordrecht, 1988).

\bibitem {Rodriguez98}C. C. Rodr\'{\i}guez, see section 3 of ``Are we cruising
a hypothesis space?'' in \emph{Maximum Entropy and Bayesian Methods}, ed. by
W. von der Linden, V. Dose, R. Fischer and R. Preuss (Kluwer, Dordrecht, 1999).

\bibitem {Rodriguez02}C. C. Rodr\'{\i}guez: `Entropic Priors for Discrete
Probabilistic Networks and for Mixtures of Gaussian Models'. In:
\emph{Bayesian Inference and Maximum Entropy Methods in Science and
Engineering}, ed. by R. L. Fry, AIP Conf. Proc. \textbf{617}, 410 (2002)
(online at arXiv.org/abs/physics/0201016).

\bibitem {Caticha00a}A. Caticha, `Maximum entropy, fluctuations and priors',
in \emph{Bayesian Methods and Maximum Entropy in Science and Engineering}, ed.
by A. Mohammad-Djafari, AIP Conf. Proc. \textbf{568}, 94 (2001) (online at arXiv.org/abs/math-ph/0008017).

\bibitem {footnote2}The number and the wording of our axioms differs from
Skilling's because we concentrate on the specific problem of ranking
probability distributions while he was concerned with ranking general positive
additive distributions. Proofs, which are easily constructed following Shore
and Johnson \cite{ShoreJohnson80} and Skilling \cite{Skilling88}, will be
presented elsewhere.

\bibitem {Amari85}S. Amari, \emph{Differential-Geometrical Methods in
Statistics} (Springer-Verlag, 1985); for a brief derivation see A. Caticha,
\textquotedblleft Change, Time and Information Geometry,\textquotedblright\ in
\emph{Bayesian Methods and Maximum Entropy in Science and Engineering}, ed. by
A. Mohammad-Djafari, AIP Conf. Proc. \textbf{568}, 72 (2001) (online at arXiv.org/abs/math-ph/0008018).

\bibitem {Skilling96}J. Skilling and S. Sibisi, \textquotedblleft Priors on
Measures\textquotedblright\ in \emph{Maximum-Entropy and Bayesian Methods}, K.
M. Hanson and R. N. Silver (eds.) (Kluwer, Dordrecht, 1996); J. Skilling,
\textquotedblleft Massive Inference and Maximum Entropy\textquotedblright\ in
\emph{Maximum-Entropy and Bayesian Methods}, G. J. Erickson, J. T. Ryckert and
C. R. Smith (eds.) (Kluwer, Dordrecht, 1998).
\end{thebibliography}
\end{document}